\documentclass[doublecol]{epl2}
\title{$s-$wave symmetry along the $c-$axis and $s+d$ in-plane superconductivity
in bulk YBa$_2$Cu$_4$O$_8$ }
 \shorttitle{$s-$wave symmetry along
the $c-$axis and $s+d$ in-plane superconductivity}

\author{R.~Khasanov\inst{1}, A.~Shengelaya\inst{2}, A.~Bussmann-Holder\inst{3},
J.~Karpinski\inst{4}, H.~Keller\inst{1}, and
K.A.~M\"uller\inst{1}} \shortauthor{R.~Khasanov\etal}

\institute{
  \inst{1} Physik-Institut der Universit\"{a}t Z\"{u}rich,
Winterthurerstrasse 190, CH-8057 Z\"urich, Switzerland \\
  \inst{2} Physics Institute of Tbilisi State University,
Chavchavadze 3, GE-0128 Tbilisi, Georgia \\
  \inst{3} Max-Planck-Institut f\"ur Festk\"orperforschung,
Heisenbergstrasse 1, D-70569 Stuttgart, Germany \\
  \inst{4} Solid State Physics Laboratory, ETH Z\"urich, CH-8093
Z\"urich, Switzerland \\
}


\abstract{ To clarify the order parameter symmetry of cuprates, the
magnetic penetration depth $\lambda$ was measured along the
crystallographic directions $a$, $b$, and $c$ in single crystals of
YBa$_2$Cu$_4$O$_8$ via muon spin rotation. {\it This method is
direct, bulk sensitive, and unambiguous.} The temperature
dependences of $\lambda_a^{-2}$ and $\lambda_b^{-2}$ exhibit an
inflection point at low temperatures as is typical for two-gap
superconductivity (TGS) with $s+d-$wave character  in the planes.
Perpendicular to the planes a pure s-wave gap is observed thereby
highlighting the important role of $c$-axis effects. We conclude
that these are generic and universal features in the bulk of
cuprates. }

\begin{document}

\maketitle

%Introduction

Two gap superconductivity (TGS) remained a theoretical issue only
for more than 20 years \cite{Suhl59,Moskalenko59,Kresin73}, even
though it seemed to be a natural and intriguing extension of BCS
theory for more complex materials. In 1980 TGS was finally observed
in Nb doped SrTiO$_3$ \cite{Binnig80} and believed to be a rare
exception in superconductors. Since the discovery of TGS in MgB$_2$
many more superconductors with TGS were found, including heavy
fermion compounds, making this feature more common than believed
early on.
In all above mentioned superconductors the combined order parameters
exhibit always the same symmetry, namely $s+s$ in Nb doped SrTiO$_3$
\cite{Binnig80} and MgB$_2$ \cite{s+s_MbB2} and $d+d$ in heavy
fermion  compounds \cite{d+d_heavy-fermion}. A totally novel
situation is met in cuprate high-temperature superconductors (HTS),
since the order parameters are of different symmetries, i.e.,  $s+d$
\cite{Muller95,MullerKeller97,Willemin98,Limonov98,Muller02,
Deutscher05,Furrer05}. However, theoretical modelling suggested a
single $d-$wave order parameter in the CuO$_2$ planes and
unfortunately biased further research and partly inhibited the
experimental efforts to characterize TGS in more detail in HTS.

Recently, new muon spin rotation ($\mu$SR) investigations of single
crystal La$_{1.83}$Sr$_{0.17}$CuO$_4$ detected an inflection point
in the temperature dependence of the inverse-squared in-plane
magnetic penetration depth $\lambda_{ab}^{-2}$, which is a direct
consequence of two superconducting gaps with largely different zero
temperature gap values \cite{Khasanov06_La214}. Since it is unclear
whether these observations are a material specific property or
generic and intrinsic to HTS, the previous $\mu$SR studies were
extended to single crystal YBa$_2$Cu$_4$O$_8$ and were performed by
applying a magnetic field along the crystallographic directions $a$,
$b$, and $c$. Thereby we obtain the three principle components of
the second moments of the local magnetic field distribution $P(B)$
in the mixed state, which are related to the superfluid density, and
reflect directly the corresponding penetration depths $\lambda_a$,
$\lambda_b$, and $\lambda_c$.
While $\lambda_a^{-2}$ and $\lambda_b^{-2}$ vary almost linearly
with temperature for $20< T<50$~K, as is expected for a $d-$wave
order parameter, the $c-$axis response ($\lambda_c^{-2}$) saturates
below 30~K, as expected for a  $s-$wave order parameter. In
addition, $\lambda_a^{-2}$ and $\lambda_b^{-2}$, both exhibit an
inflection point in their temperature dependences around
$T\simeq$10~K which -- as has been shown in \cite{Bussmann-Holder06}
-- is the consequence  of two coexisting order parameters, namely
$s+d$.

%Experimental
Details of the sample preparation for YBa$_2$Cu$_4$O$_8$ can be
found elsewhere \cite{Y124_sample_preparation}.  All crystals used
in the present study were taken from one batch.
The superconducting transition temperature $T_c$ and the width of
the superconducting transition $\Delta T_c$ were determined for the
three sets ($\simeq$40-50 crystals each) of the main set
($\simeq$130) of crystals. Both were obtained from field-cooled (0.5
mT) magnetization curves measured by a SQUID magnetometer and
exhibited all the same values, i.e., $T_c\simeq79.9$~K and $\Delta
T_c\simeq2$~K.
The crystals had mostly a rectangular shape with a typical size of
approximately 0.8x0.3x0.05~mm$^3$. X-ray measurements revealed that
the crystallographic $b-$axis is exactly parallel to the longest
side. Bearing in mind that the $c-$axis is perpendicular to the flat
surface of the crystal, we were able to orient the whole set  along
the crystallographic $a$, $b$, and $c$ directions. The final
orientation of the crystals in the mosaic was checked by using a
polarizing microscope.

\begin{figure}[htb]
%\centering
\onefigure[width=0.9\linewidth]{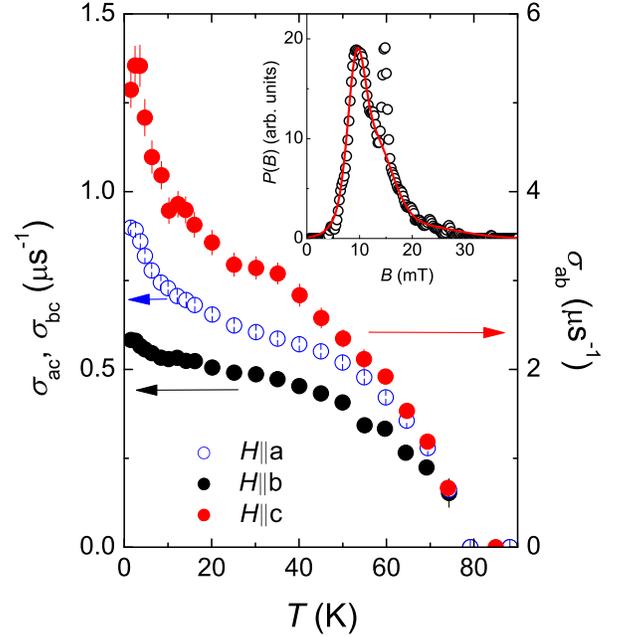}
%\includegraphics[width=0.8\linewidth]{Figure1}
% \vspace{-1.0cm}
%
\caption{(Color online) Temperature dependences of
$\sigma_{ab}\propto \lambda^{-2}_{ab}$ ($H\|c$), $\sigma_{ac}\propto
\lambda^{-2}_{ac}$ ($H\|b$), and $\sigma_{bc}\propto
\lambda^{-2}_{bc}$ ($H\|a$) of YBa$_2$Cu$_4$O$_8$, measured after
field cooling the sample in $\mu_0H=$0.015~T. The inset shows the
local magnetic field distribution $P(B)$ obtained by means of the
maximum entropy Fourier transform technique at $T=1.7$~K and
$\mu_0H=0.015$~T applied parallel to the $c-$axis (open circles:
data; solid line: four component Gaussian fit excluding the
background).}
 \label{fig:sigma_ij}
\end{figure}

The transverse-field $\mu$SR experiments on a mosaic of oriented
YBa$_2$Cu$_4$O$_8$ single crystals  were done at the $\pi$M3 and
$\pi$E1 beam lines at the Paul Scherrer Institute (Villigen,
Switzerland). The mosaic was field cooled from above $T_c$ to 1.7~K
in a field of 0.015~T. The $\mu$SR experiments were performed for
the magnetic field applied parallel to the $a$, $b$, and $c$
crystallographic axes. Typical counting statistics were $\sim$24-30
million muon detections over three detectors. In the analysis
presented below we used the well-known fact that for an extreme
type-II superconductor in the mixed state $\lambda^{-4}$ is
proportional to the second moment of $P(B)$ probed by $\mu$SR
\cite{Brandt88}. The second moment of $P(B)$ was calculated within
the same framework described in \cite{Khasanov06_La214}. We used a
four component Gaussian expression to fit the $\mu$SR time spectra.
One component arises here from the background signal stemming from
muons stopped outside the sample, whereas the other three components
describe the asymmetric line shape of $P(B)$ in the mixed state (see
inset to Fig.~\ref{fig:sigma_ij}). The first and the second moments
of $P(B)$ (excluding the background component) are then obtained as:
%\cite{Khasanov06_La214}:
%
\begin{equation}
\langle B \rangle=\sum_{i=1}^3{A_i B_i \over A_1+A_2+A_3} \,
\label{eq:B_mean}
\end{equation}
and
\begin{equation}
\langle \Delta B^2
\rangle=\frac{\sigma^2}{\gamma^2_\mu}=\sum_{i=1}^3{A_i \over
A_1+A_2+A_3} \left[ \frac{\sigma_i^2}{\gamma_{\mu}^2} +[B_i-
\langle B \rangle]^2 \right] .
\label{eq:dB}
\end{equation}
Here $\gamma_{\mu} = 2\pi\times135.5342$~MHz/T is the muon
gyromagnetic ratio. $A_i$, $\sigma_i$, and $B_i$ are the asymmetry,
the relaxation rate, and the first moment of the $i-$th component,
respectively.
The superconducting part of the square root of the second moment
($\sigma_{sc}\propto\lambda^{-2}$) was then obtained by subtracting
the nuclear moment contribution ($\sigma_{nm}$) measured at $T>T_c$,
according to $\sigma_{sc}^2=\sigma^2 - \sigma_{nm}^2$.
%\cite{Khasanov06_La214}.
%
To ensure that the increase of the second moment of the measured
$\mu$SR signal below $T_c$ is attributed entirely to the vortex
lattice, zero-field $\mu$SR experiments were performed. No evidence
for static magnetism in the YBa$_2$Cu$_4$O$_8$ mosaic sample down to
1.7~K was observed.

For an anisotropic London superconductor the effective penetration
depth for the magnetic field along the $i$-th crystallographic
axis is given by \cite{Ager00}:
\begin{equation}
\lambda_{ jk}^{-2}=\frac{1}{\lambda_j\lambda_k} \propto\sigma_{jk}.
 \label{eq:lambda_jk}
\end{equation}
Here the index ''$sc$`` for the superconducting part of the square
root of the second moment $\sigma_{sc}$ is omitted for simplicity.
From Eq.~(\ref{eq:lambda_jk}) it is obvious that for the magnetic
field applied along one of the principal axes $a$, $b$, and $c$, the
components $\sigma_{bc}\propto\lambda_{bc}^{-2}$,
$\sigma_{ac}\propto\lambda_{ac}^{-2}$, and
$\sigma_{ab}\propto\lambda_{ab}^{-2}$ are measured. The temperature
dependences of $\sigma_{bc}$, $\sigma_{ac}$, and $\sigma_{ab}$ after
field-cooling the sample in $\mu_0 H=0.015$~T are shown in
Fig.~\ref{fig:sigma_ij}. It is seen that at $T_{ip}\sim 10-20$~K all
measured $\sigma_{ij}(T)$ exhibit an inflection point. Below this
point $\sigma_{ab}$, $\sigma_{bc}$, and $\sigma_{ac}$ increase by
approximately 70\%, 35\%, and 10\%, respectively. Note, that a
similar inflection point in $\lambda_{ab}^{-2}(T)$ was also observed
in low-field magnetization (LFM) experiments on powder samples of
YBa$_2$Cu$_4$O$_8$ \cite{Panagopoulos99}. However, in these
experiments the increase of $\lambda_{ab}^{-2}$ below $T_{ip}$ was
much less pronounced than the one observed in the present study.
This difference can be explained by the fact that LFM probes the
penetration depth mainly near the surface, whereas $\mu$SR measures
$\lambda$ in the {\it bulk}. In LFM experiments the magnetic field
penetrates the sample only on a distance $\lambda$ from the surface
of the sample (few hundred nanometers), thereby leaving the main
part of the superconducting volume unaffected. In contrast, muons
penetrate at much longer distances into the sample (few-tenth of a
millimeter), thus probing $\lambda$ deeply inside the sample. It
should be recalled that at the surface due to the symmetry breaking,
the order parameter has pure $d-$wave character, whereas in the bulk
$s+d-$wave symmetry is present \cite{Muller02}. This explains in a
natural way the difference between the surface sensitive LFM and the
bulk sensitive $\mu$SR experiments.

\begin{figure}[htb]
%\centering
\onefigure[width=0.9\linewidth]{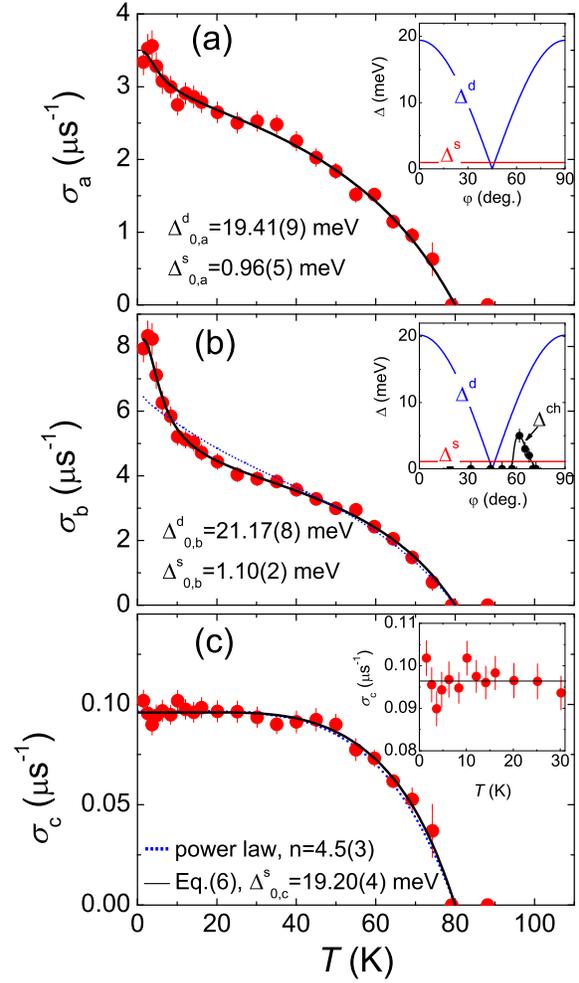}
%\includegraphics[width=1.0\linewidth]{Figure2}
%\includegraphics[width=0.7\linewidth]{Figure2}
% \vspace{-1.0cm}
%
\caption{(Color online) Temperature dependences of
$\sigma_{a}\propto \lambda^{-2}_{a}$ (a), $\sigma_{b}\propto
\lambda^{-2}_{b}$ (b), and $\sigma_{c}\propto \lambda^{-2}_{c}$ (c)
of YBa$_2$Cu$_4$O$_8$ at $\mu_0H=$0.015~T obtained from the data
presented in Fig.~\ref{fig:sigma_ij} by means of
Eq.~(\ref{eq:lambda_i}). Lines in (a) and (b) represent results of
the fits of $\sigma_a(T)$ and $\sigma_{b}(T)$ by means of the
two-component [Eq.~(\ref{eq:sigma_two-gap})]  and the
three-component models. The blue dotted curve in (c) represents a
fit to the data assuming that $\sigma_b(T)$ is entirely determined
by the in-plane $d-$wave and the chain gap measured in
\cite{Khasanov06_ARPES} (see text for details). The black solid line
and the blue dotted curve in (c) represent fits by means of
Eq.~(\ref{eq:sigma-d}) with $g^s(\varphi)=1$ and the power law
$\sigma_c(T)=\sigma_c(0)[1-(T/T_c)^n]$, respectively. The insets in
(a) and (b) represent the angular dependences of the $d-$wave gap
($\Delta^d$), the $s-$wave gap ($\Delta^s$) and the gap in CuO
chains ($\Delta^{ch}$) obtained in \cite{Khasanov06_ARPES}. The
inset in (c) shows $\sigma_{c}(T)$ at $T<30$K. }
 \label{fig:sigma_i}
\end{figure}

From Eq.~(\ref{eq:lambda_jk}) the individual components of $\lambda$
along the $i-$th crystallographic direction are obtained as:
\begin{equation}
\sigma_i=\frac{\sigma_{ij}\sigma_{ik}}{\sigma_{jk}}
\propto\frac{1}{\lambda_i^2}.
 \label{eq:lambda_i}
\end{equation}
Figure~\ref{fig:sigma_i} shows the temperature dependences of
$\sigma_a\propto\lambda_a^{-2}$, $\sigma_b\propto\lambda_b^{-2}$,
and $\sigma_c\propto\lambda_c^{-2}$ as derived from the data
presented in Fig.~\ref{fig:sigma_ij} by using
Eq.~(\ref{eq:lambda_i}). In the following we discuss the
temperature dependence of each particular component separately.

From Figs.~\ref{fig:sigma_i}~(a) and (b) it is evident that both
$\sigma_{a}\propto\lambda_{a}^{-2}$ and
$\sigma_{b}\propto\lambda_{b}^{-2}$ increase almost linearly with
decreasing temperature in the range $50 < T< 20$~K. Around
$T_{ip}\simeq10$~K an inflection point is visible in the temperature
dependence.
Below this point an unusual increase in both quantities appears:
$\sigma_{a}$ increases by almost 20\% (from 2.95~$\mu$s$^{-1}$ at
$T=10$~K to 3.5~$\mu$s$^{-1}$ at $T=2.6$~K) and $\sigma_{b}$ by 60\%
(from 5.2~$\mu$s$^{-1}$ at $T=10$~K to 8.3~$\mu$s$^{-1}$ at
$T=2.6$~K). As has been shown in Ref.~\cite{Bussmann-Holder06}, an
inflection point in $\lambda^{-2}(T)$ suggests the presence of at
least two superconducting gaps in YBa$_2$Cu$_4$O$_8$ with very
different gap values, i.e., a large gap and a small one. The same
behavior was observed recently in La$_{1.83}$Sr$_{0.17}$CuO$_4$ by
$\mu$SR \cite{Khasanov06_La214}, as well as in other HTS using
various techniques
\cite{Willemin98,Limonov98,Muller02,Deutscher05,Furrer05},
supporting a two-gap behavior with the larger gap being of $d-$wave
and the smaller one of $s-$wave symmetry. The above data were
analyzed by assuming that an $s-$wave and a $d-$wave gap contribute
to $\sigma$ according to:
% \cite{Khasanov06_La214}:
%
\begin{equation}
\sigma(T)=\sigma^{d}(T)+\sigma^{s}(T),
 \label{eq:sigma_two-gap}
\end{equation}
where both components are expressed like \cite{Khasanov06_La214}:
\begin{equation}
\frac{\sigma(T,\Delta_0)}{\sigma(0)}=  1+ \frac{1}{\pi}
\int_{0}^{2\pi}\int_{\Delta(T,\varphi)}^{\infty}\left(\frac{\partial
f}{\partial E}\right)\frac{E\
dEd\varphi}{\sqrt{E^2-\Delta(T,\varphi)^2}}\ .
 \label{eq:sigma-d}
\end{equation}
Here, $f=[1+\exp(E/k_BT)]^{-1}$ is  the Fermi function, $\Delta_0$
is the maximum value of the gap, and  $\Delta(T,\varphi)=\Delta_0
\tilde{\Delta}(T/T_c)g(\varphi)$. For the normalized gap
$\tilde{\Delta}(T/T_c)$ tabulated values of
Ref.~\cite{Muhlschlegel59} were used. The function $g(\varphi)$
describes the angular dependence of the gap and is given by
$g^d(\varphi)=|\cos(2\varphi)|$ for the $d-$wave gap
\cite{Deutscher05} and $g^s(\varphi)=1$ for the $s-$wave gap [see
insets in Figs.~\ref{fig:sigma_i}~(a) and (b)]. From this analysis
we obtain $\Delta^d_{0,a}=19.41(9)$~meV,
$\sigma_a^d=3.03(2)$~$\mu$s$^{-1}$, $\Delta^s_{0,a}=0.96(5)$~meV,
$\sigma_a^s=0.48(2)$~$\mu$s$^{-1}$ and
$\Delta^d_{0,b}=21.17(8)$~meV, $\sigma_b^d=4.66(2)$~$\mu$s$^{-1}$,
$\Delta^s_{0,b}=1.10(2)$~meV, $\sigma_b^s=3.63(2)$~$\mu$s$^{-1}$ for
$\sigma_a(T)$ and $\sigma_b(T)$, respectively. The main results are
summarized in Table~\ref{Table:two-gap}.

\begin{table}[h]
\caption[~]{\label{Table:two-gap} Summary of the gap analysis of the
temperature dependences of $\sigma_{i}\propto \lambda^{-2}_{i}$
($i=a$, $b$, $c$) for YBa$_{2}$Cu$_{4}$O$_8$. The meaning of the
parameters is explained in the text.}
\begin{center}
 \vspace{-0.7cm}
\begin{tabular}{ccccccccc}\\ \hline
\hline
 Component&$\Delta^d_{0,i}$&$\Delta^s_{0,i}$&
 $\gamma_{ab}^{\Delta^d}$&$\gamma_{ab}^{\Delta^s}$\\
&(meV)&(meV)&&\\
\hline
$\sigma_a$&19.41(9)&0.96(5)&--&--\\
$\sigma_b$&21.17(8)&1.10(2)&1.09(1)&1.15(6)\\
$\sigma_c$&--&19.20(4)&--&--\\
 \hline
 \hline
\end{tabular}
   \end{center}
\end{table}

The temperature dependence of $\sigma_{c}\propto\lambda^{-2}_c$
differs significantly from the one of the in-plane components since
it saturates at temperatures $T<30K$ to become $T$ independent [see
inset in Fig.~\ref{fig:sigma_i}~(c)]. This dependence is analyzed in
two ways: (i) by using Eq.~(\ref{eq:sigma-d}) with $g^s(\varphi)=1$,
yielding $\Delta^s_{0,c}=19.20(4)$~meV, and (ii) by assuming the
phenomenological power law $\sigma_c(T)=\sigma_c(0)[1-(T/T_c)^n]$.
The results are given in Table~\ref{Table:two-gap} and compared to
the experimental data in Fig.~\ref{fig:sigma_i}~(c), from which it
is obvious that both approaches are almost undistinguishable. For
the power law dependence a critical exponent $n=4.5(3)$ is derived
which is close to the one obtained within a two-fluid model where
$n=4$, which applies to a strong coupling $s-$wave BCS
superconductors \cite{Rammer88}. Using a $d-$wave model for a clean
superconductor, Ref.~\cite{Xiang96} predicts an exponent $n=5$.
Since the observed power law exponent lies between these two values,
it could be argued that $d-$wave superconductivity could also be the
cause of the temperature dependence of
$\sigma_c\propto\lambda_c^{-2}$. However, this can be ruled out,
since the results of Ref.~\cite{Xiang96} do not apply to cuprate
superconductors containing chains, like YBa$_2$Cu$_3$O$_{7-\delta}$
and YBa$_2$Cu$_4$O$_8$. We can thus safely conclude that the order
parameter along the $c-$axis is isotropic $s-$wave. This conclusion
is further supported by $c-$axis tunneling \cite{Sun94}, bicristal
twist Josephson junctions \cite{Li99}, optical pulsed probe
\cite{Kabanov99}, and optical reflectivity (see Fig.~4 in
Ref.~\cite{Muller03}) experiments, as well as theoretical
considerations \cite{Klemm00}. It is important to note that due to
the reasons described above the saturation of $\lambda_c^{-2}(T)$
obtained here has not been observed in LFM experiments
\cite{Panagopoulos99}.

The anisotropy observed in $\sigma_a$ and $\sigma_b$  and their
unconventional temperature dependences deserve further remarks. The
temperature dependences of both quantities are characterized by an
inflection point at $T_{ip}\simeq10$~K evidencing that at least two
superconducting gaps contribute to the superfluid density. However,
along the crystallographic $b-$direction this is much more
pronounced than along $a$, and the analysis of both data sets in
terms of $s+d$ wave gaps using Eq.~(\ref{eq:sigma_two-gap}) reveals
that along the $a$ direction the $s-$wave order parameter
contributes $14$\%, whereas along the $b$ direction the $s-$wave
contribution is 42\%. This observation is a consequence of the
structural anisotropy caused by the CuO chains constitutes itself by
an enhanced $s-$wave superconducting density along the chain
direction which can have various origins. Early on, Raman
experiments have detected the opening of a second gap at low
frequencies which appears only in specific scattering geometries
\cite{Heyen91}. Since these observations have only been made in the
YBCO family, they were interpreted as being due to the opening of a
gap in the chains. In addition, recent angle resolved photoemission
data (ARPES) have observed a coherence peak related to the chain
bands, which appears in a very limited $k-$space region
\cite{Khasanov06_ARPES}. It was found that in YBa$_2$Cu$_4$O$_8$ the
superconducting gap in the chains opens only at certain angles
$\varphi$ and reaches its maximum value of 5(1)~meV at
$\varphi\simeq60^\circ$ [see inset in Fig.~\ref{fig:sigma_i}~(b)].
Therefore, one may speculate that the two-gap behavior in
YBa$_2$Cu$_4$O$_8$ is entirely determined by the in-plane $d-$wave
and the chain gap. This scenario can, however, be excluded.  First
of all, the fit of Eq.~(\ref{eq:sigma_two-gap}) to $\sigma_b(T)$
taking into account a chain gap of this symmetry, leads to a rather
poor agreement with the experimental data [blue dotted curve in
Fig.~\ref{fig:sigma_i}~(b)].
Second, the observation of an inflection point in $\sigma_a(T)$
cannot be explained by a possible misalignment of the crystals in
the mosaic. $\sigma_{a}^s(0)=0.48$~$\mu s^{-1}$ would correspond to
a situation when {\it all} the crystals in the mosaic are turned by
$\pm$8$^\circ$, or $\simeq$30$^\circ$ FWHM if one assumes a
triangular distribution of orientations with the maximum along the
crystallographic $b$ axis.
Therefore, the data presented in Fig.~\ref{fig:sigma_i}~(b) were
analyzed by assuming that $\sigma_{b}(T)$ is the sum of three
components: $\sigma^d_b(T)$, $\sigma^s_b(T)$, and the contribution
from the chains $\sigma^{ch}_b(T)$. These results are, however,
undistinguishable from the $s+d$ analysis presented above.
It is important to note, that independent of a possible chain
related energy gap, both procedures provide clear evidence that the
in-plane superconducting order parameter consists of two components,
a $d-$wave order parameter and an additional $s-$wave component.

The maximum values of the in-plane $d-$wave gap along the $a$ and
$b$ directions are of similar order of magnitudes, i.e.,
$\Delta^{d}_{0,a}=19.41(9)$~meV, $\Delta^{d}_{0,b}=21.17(8)$~meV and
thus in a good agreement with $\Delta^d_0=22$~meV derived from
tunnelling experiment \cite{Karpinski96}. The in-plane anisotropy in
both gap values, $s$ and $d$, is approximately the same, i.e.,
$\gamma_{ab}^{\Delta^d}=\Delta^{d}_{0,b}/\Delta^{d}_{0,a}=1.09(1)$
and
$\gamma_{ab}^{\Delta^s}=\Delta^{s}_{0,b}/\Delta^{s}_{0,a}=1.15(6)$
(see Table~\ref{Table:two-gap}). Moreover, it is important to note
that the $s-$wave gap along the $c$ direction
$\Delta^{s}_{0,c}=19.20(4)$~meV is of the same order of magnitude as
the $d-$wave gaps in the $ab$ plane: $\Delta^{s}_{0,c}
\simeq\Delta^{d}_{0,a} \simeq\Delta^{d}_{0,b} \approx 20$~meV.

In conclusion, we performed a systematic $\mu$SR study of the
magnetic penetration depths $\lambda_a$, $\lambda_b$, and
$\lambda_c$ on single crystals of YBa$_2$Cu$_4$O$_8$. In contrast to
previous LFM experiments \cite{Panagopoulos99}, our method is {\it
bulk} sensitive and a direct probe of the penetration depth. The use
of single crystals enables us to derive  the  magnetic penetration
depth along the three principal crystallographic directions. Along
the $a$ and $b$ directions clear evidence is obtained that TGS is
realized also in the chain containing compound YBa$_2$Cu$_4$O$_8$.
While the in-plane penetration depth is anisotropic, exhibiting an
inflection point at low temperatures in both $\lambda_a^{-2}$ and
$\lambda_b^{-2}$ which is characteristic of TGS, the $c-$axis data
provide clear evidence for an  an isotropic $s-$wave gap. From the
data it must be concluded that the in-plane superconducting gap
consists of two components, namely $s+d$. The situation along the
$c-$axis is different and supports $s-$wave only.
This exceptional behavior has not been predicted by any theory,
since the third dimension has mostly been neglected.
Since YBa$_2$Cu$_4$O$_8$ differs structurally substantially from the
previously investigated system La$_{2-x}$Sr$_x$CuO$_4$, where the
in-plane penetration depth also shows $s+d$ wave superconductivity
\cite{Khasanov06_La214}, the above findings cannot be attributed to
specific structural features of the chain containing compound, but
show that $s+d$ order parameters are  generic,  intrinsic and common
to all HTS. The results thus exclude theoretical approaches as e.g.,
the ''plain vanilla`` mechanism which concentrates on the CuO$_2$
plane only. Besides of the fact that t-J or 2D Hubbard models are
not capable to yield the observed $s-$wave component, the role of
the lattice played for superconductivity has to be reconsidered
since the presence of an $s-$wave component naturally points to its
importance.

This work was partly performed at the Swiss Muon Source (S$\mu$S),
Paul Scherrer Institute (PSI, Switzerland). The authors are grateful
to A.~Amato and R.~Scheuermann for assistance during the $\mu$SR
measurements. This work was supported by the Swiss National Science
Foundation, the K.~Alex~M\"uller Foundation, and in part by the
SCOPES grant No. IB7420-110784, the EU Project CoMePhS, and the NCCR
program MaNEP.

\end{document}